\documentclass[prl,amsmath,amssymb,twocolumn,showpacs,floatfix,amsfonts]{revtex4}
\usepackage{graphicx}
\def \beq {\begin{equation}}
\def \eeq {\end{equation}}

\begin{document}
\title{Sub-Shot-Noise Magnetometry with a Correlated Spin-Relaxation Dominated Alkali-Metal Vapor}
\author{I. K. Kominis}
\email{ikominis@iesl.forth.gr}

\affiliation{Department of Physics, University of Crete, Heraklion 71103, Greece}
\affiliation{Institute of Electronic Structure and Laser, Foundation for Research and Technology,Heraklion 71110, Greece}

\date{\today}

\begin{abstract}
Spin noise sets fundamental limits to the precision of
measurements using spin-polarized atomic vapors, such as performed
with sensitive atomic magnetometers. Spin squeezing offers the
possibility to extend the measurement precision beyond the
standard quantum limit of uncorrelated atoms. Contrary to the
current understanding, we show that even in the presence of spin
relaxation, spin squeezing can lead to a significant reduction of
spin noise, and hence an increase in magnetometric sensitivity,
for a long measurement time. This is the case when correlated spin
relaxation due to binary alkali-atom collisions dominates
independently acting decoherence processes.
\end{abstract}
 \maketitle Quantum noise due
to the fundamental quantum-mechanical uncertainties of physical
observables sets the standards quantum limits \cite{braginsky}
(SQL) for the accuracy of any quantum measurement. Spin-projection
noise or spin noise \cite{projection_noise}, in particular, poses
a fundamental limit to the measurement precision using an ensemble
of spin systems, be it the actual spin angular momentum of
alkali-metal atoms employed, for example, in sensitive
magnetometers \cite{romalis2002}, or other two-level systems such
as those involved in atomic clocks \cite{santarelli}. Spin noise
of an ensemble of uncorrelated atoms leads to a fundamental noise
level that scales as 1/$\sqrt{N}$, where $N$ is the number of
atoms participating  in the measurement process. The creation of
quantum correlations between the atoms has emerged as a
possibility of extending the measurement precision beyond the SQL
of uncorrelated ensembles. Spin squeezing refers to multi-particle
quantum states of the system in consideration which exhibit this
suppression of quantum noise in spectroscopic measurements
\cite{wineland}. Several theoretical proposals describing ways to
create spin squeezing have appeared \cite{theory}, but so far,
spin squeezing has been
 experimentally demonstrated in systems where decoherence is negligible, i.e. in a cold cesium vapor \cite{mabuchi,polzik_prl} and in a low
 density thermal cesium vapor \cite{bigelow_prl}.
The motivation for this work is the possibility of enhancing the magnetic sensitivity of atomic mangetometers \cite{nature_magnetometer}
 employing high density alkali-metal vapors by creating spin-squeezed states. These devices have several applications \cite{appl,meg}
 which would benefit from an increased sensitivity beyond the
relevant SQL.\newline\indent
 However, it was recently conjectured \cite{budker_qnd}
 that spin squeezing is of little use in the presence of spin relaxation, leading to sub-SQL magnetic sensitivities only for an impractically
 short
 measurement time. This would be detrimental since, unlike laser-cooled atomic ensembles, in high density thermal atomic vapors used in atomic magnetometers spin relaxation is a
 dominant effect. A similar result was derived for the case of improving frequency standards by use of entanglement \cite{ekert}.\newline\indent
In this Letter we show that spin squeezing does actually lead to a
sub-SQL spin noise level and enhanced magnetic sensitivity even in
the presence of spin relaxation, and for long measurement times.
Using quantum state diffusion theory \cite{gisin}, which naturally
reflects the fluctuation-dissipation theorem for the collective
atomic spin of uncorrelated atoms, we demonstrate the intimate
connection between spin noise and spin relaxation. Hence we find
that there is no additional noise due to spin relaxation as
suggested in \cite{budker_qnd}, where spin relaxation was treated
independently of spin noise. We then show that the dominant
relaxation mechanism in a dense alkali-metal vapor, i.e. binary
alkali-metal atom collisions, preserves the ensemble quantum
correlations, allowing an enhanced measurement precision for a
time on the order of the spin relaxation time. We also identify
the opposite limit, in which independently acting decoherence
mechanisms, if dominant, do actually lead to the conclusions
reported in \cite{budker_qnd,ekert}.
\newline\indent
The physical system we will be considering is a thermal ensemble
of alkali-metal atoms confined in a cell. The atoms are initially
spin-polarized along the $\mathbf{\hat{x}}$-axis, so that $\langle
s_{x}\rangle=1/2$, where $\mathbf{s}$ denotes the atom's electron
spin.
%\begin{figure}[htp]
%\includegraphics[width=5cm]{eps_files/schematic.eps}
%\caption{\label{fig:schematic}(Color Online) Typical scheme of atomic magnetometer. PD stands for photodiode.}
%\end{figure}
A small magnetic field to be measured is applied along the
$\mathbf{\hat{z}}$-axis, and induces a precession of the spins,
observed for a measurement time $\tau$. The transverse
spin-polarization thus produced can be detected \cite{budker_rmp}
via Faraday rotation of an off resonant probe laser's polarization
measured with, e.g., a balanced polarimeter. In a dense
alkali-metal vapor the transverse spin relaxation, or spin
decoherence, is dominated by two kinds of binary collisions,
namely spin-exchange and spin-destruction collisions
\cite{appelt_pra}, with respective rates $1/T_{\rm se}$ and
$1/T_{\rm sd}$, proportional to the atom density. Both are
"sudden" with respect to the nuclear spin and tend to reduce the
density matrix \cite{happer_rmp}
$\rho=\phi+\mathbf{a}\cdot\mathbf{s}$ (with $\phi$ and
$\mathbf{a}$ being nuclear operators) to the part $\phi$ without
electronic spin-polarization, i.e. $d\rho/dt=(\phi-\rho)/T_{2}$,
where $1/T_{2}=1/T_{\rm se}+1/T_{\rm sd}$. At very low magnetic
field, which will be assumed henceforth, relaxation due to
spin-exchange is suppressed \cite{happer_tang}, and
$1/T_{2}=1/T_{\rm sd}$. The time evolution of the density matrix
can be written in the Lindblad form \beq {{d\rho}\over
{dt}}=-i[{\cal H}_{g},\rho]+\sum_{j=1}^{3}{(L_{j}\rho
L_{j}^{\dagger}-{1\over 2}L_{j}^{\dagger}L_{j}\rho- {1\over 2}\rho
L_{j}^{\dagger}L_{j})} \eeq where ${\cal H}_{g}$ is the ground
state hamiltonian and the three Lindblad operators are \beq
L_{1}={1\over \sqrt{2T_{2}}}s_{+},~~~L_{2}=
{1\over\sqrt{2T_{2}}}s_{-},~~~L_{3}={1\over \sqrt{T_{2}}}s_{z}
\eeq It can be shown \cite{percival_book} that the change of a
quantum state $|\psi\rangle$ after the elapse of a time interval
$dt$ is given by the quantum state diffusion equation
\begin{align}
|d\psi\rangle&=-idt{\cal H}_{g}|\psi\rangle\nonumber  \\
&+\sum_{j=1}^{3}{\left(\langle L_{j}^{\dagger}\rangle_{\psi} L_{j}- {1\over 2}\langle L_{j}^{\dagger}\rangle_{\psi}\langle
L_{j}\rangle_{\psi}-{1\over 2}L_{j}^{\dagger}L_{j}
\right)}|\psi\rangle dt\nonumber  \\
&+\sum_{j=1}^{3}{(L_{j}-\langle L_{j}\rangle_{\psi})|\psi\rangle d\eta_{j}} \label{eq:dpsi}
\end{align}
where the first term represents the hamiltonian evolution, the
second dissipation, and the third the stochastic fluctuations,
described by the statistically independent complex Wiener
processes $d\eta_{j}$, with $j$=1,2,3, i.e.
$d\eta_{i}d\eta_{j}^{*}=dt\delta_{ij}$. Since the nuclear spin
plays no fundamental role in the following considerations, we will
consider an ensemble of $N$ spin-1/2 particles. We furthermore
assume the the probing laser is far enough off resonance that we
can neglect measurement-induced back-action on the spins
\cite{geremia_pra}, that is, we are going to only consider
spontaneous spin noise and its effect on measurement precision.
This can be done since the probe polarization rotation scales with
the probe laser detuning $\Delta$ as $1/\Delta$, whereas the
measurement strength \cite{geremia_pra} scales as $1/\Delta^{2}$.
The effects of probe photon shot-noise and scattering have been
treated in \cite{budker_qnd}. From (\ref{eq:dpsi}) it follows
\cite{katso} that the expectation value $\langle s_{y}\rangle$,
which is the measured observable, obeys an Ornstein-Uhlenbeck
stochastic process, \beq d\langle s_{y}\rangle=\omega_{L}\langle
s_{x}\rangle_{0}dt-{{dt}\over {2T_{2}}}\langle
s_{y}\rangle+{d\eta\over \sqrt{4T_{2}}},\label{eq:spin_fl} \eeq
where $\omega_{L}$ is the Larmor frequency and now $d\eta$ is a
real Wiener process \cite{gillespie}, i.e. a normal random
variable with zero mean and variance $dt$. Defining the ensemble
transverse spin as $S_{y}=\sum_{i=1}^{N}{s_{y}^{(i)}}$, it follows
that for $N$ uncorrelated atoms \beq d\langle
S_{y}\rangle=\omega_{L}\langle S_{x}\rangle_{0}dt-{{dt}\over
{2T_{2}}}\langle S_{y}\rangle+\Delta S_{y}{{d\eta}\over
\sqrt{T_{2}}},\label{eq:total_spin_fl} \eeq where $\langle
S_{x}\rangle_{0}=N\langle s_{x}\rangle_{0}$ and $\Delta
S_{y}=\sqrt{N/4}$ is the coherent spin state (CSS) uncertainty.
The spectrum of the transverse spin fluctuations follows a
Lorentzian distribution centered at the origin (at zero magnetic
field) with a width equal to $1/2T_{2}$, which also sets the
bandwidth of the magnetometric measurement \cite{katso_prl}. As is
evident from (\ref{eq:total_spin_fl}), transverse spin dissipation
and spin noise are intimately related, both being described by one
and the same parameter, $T_2$. The reason that spin noise sets the
SQL for a magnetic field measurement using a collision-dominated
alkali-metal vapor is that atomic collisions and the associated
relaxation continuously redistribute the variance of the ensemble
transverse spin, i.e. even in the infinite time limit when any
initial nonzero expectation $\langle S_{y}\rangle$ has decayed
away, $\langle S_{y}\rangle$ has a non-zero power spectrum
extending to $1/2T_{2}$. This forms the basis of spontaneous spin
noise spectroscopy \cite{nature_spin_noise,katso}. In contrast, in
the case of laser-cooled collision-less atomic vapors
\cite{mabuchi}, the power spectrum of $\langle S_{y}\rangle$ has
only a zero-frequency component, which exhibits a shot-to-shot
distribution around zero with an uncertainty $\sqrt{N/4}$
characterizing the CSS \cite{mabuchi}. We now assume that the
atomic ensemble has been spin-squeezed, i.e. the spin-squeezing
parameter \cite{projection_noise} $\xi=\sqrt{N}\Delta
S_{y}/\langle S_{x}\rangle<1$. That is, as in \cite{kitagawa}, we
assume that
 a spin squeezing
hamiltonian has been applied before probing the precessing spins.
In the presence of spin squeezing, however, the spin uncertainties
of individual atoms do not simply add in quadrature, as there are
negative pairwise correlations \cite{kitagawa} that have to be
accounted for: \beq C_{yy}^{ij}\equiv\langle
s_{y}^{(i)}s_{y}^{(j)}\rangle-\langle s_{y}^{(i)}\rangle\langle
s_{y}^{(j)}\rangle={{\xi^{2}-1}\over {4N}}, \label{eq:corr}\eeq
leading again to Eq. (\ref{eq:total_spin_fl}), but now \beq \Delta
S_{y}=\xi\sqrt{N/4}\label{eq:reduced_spin_fluct} \eeq This is the
first main result of this work. Spin relaxation obviously leads to
dissipation of a non-zero expectation value $\langle
S_{y}\rangle$. At the same time it is manifested through spin
noise, i.e. the fluctuations of $\langle S_{y}\rangle$ around its
mean value, described by the third, stochastic term of Eq.
(\ref{eq:total_spin_fl}). There is no other noise source due to
spin relaxation. We emphasize that these fluctuations are driven
by atomic collisions and not by photon noise coupling into
$\langle S_{y}\rangle$, as is the
 case with strong measurements \cite{geremia_prl} of
the collective spin of laser-cooled vapors. The magnitude of these
fluctuations, given by Eq. (\ref{eq:reduced_spin_fluct}) is indeed
reduced if $\xi<1$ and a sub-SQL sensitivity can be achieved for a
measurement time $\tau\approx T_{2}$. Indeed, from
(\ref{eq:total_spin_fl}) it follows that the maximum value of
$\langle S_{y}\rangle$ will be $\langle
S_{y}\rangle_{\tau}=\omega_{L}2T_{2}\langle S_{x}\rangle_{0}$ at
$\tau\approx T_{2}$. The noise acquired at time $\tau=T_2$ due to
the stochastic term of the evolution equation
(\ref{eq:total_spin_fl}) will on the average be $\delta\langle
S_{y}\rangle=\xi\sqrt{{ N}\over
{4T_{2}}}\sqrt{T_{2}}=\xi\sqrt{N}/2$. Thus the sensitivity limit
for measuring a small frequency $\omega_{L}$ will be (neglecting
factors of 2) \beq \delta\omega_{L}={{\delta\langle
S_{y}\rangle}\over {\partial\delta\langle
S_{y}\rangle/\partial\omega_{L}}}={\xi\over {T_{2}\sqrt{N}}} \eeq
By averaging $n$ such measurements for a total measurement time
$T=nT_{2}$, we get \beq
\delta\omega_{L}={\xi\over{\sqrt{NT_{2}T}}}\label{eq:domega} \eeq
For the case of uncorrelated atoms, $\xi=1$, we recover the
well-known shot-noise limit. In the above derivation we have
neglected the decay and fluctuations of $\langle S_{x}\rangle$.
This is allowed, since the decay of $\langle S_{x}\rangle$ during
the measurement time $T_2$ will change our estimate by a factor of
order 1, and the fluctuations of $\langle S_{x}\rangle$, of order
$\sqrt{N}$, leak into $\langle S_{y}\rangle$ but are diminished by
the spin rotation angle, $\omega_{L}T_{2}$. For a magnetic field
magnitude in the fT range and a spin coherence time $T_{2}\approx
1$ ms, this factor is of order $10^{-8}$, and for all practical
purposes, is negligible compared to the actual spin noise of
$\langle S_{y}\rangle$, of magnitude $\xi\sqrt{N}$.
\newline\indent In the following we are going to
justify the previous assertions in more detail. The reason that
spin noise is reduced even in the presence of relaxation is that
the decoherence mechanism we are considering is not acting on
every atom independently, resulting in a change $d(\Delta
S_{y})^{2}=0$ to first order in $dt$. Indeed, in a binary
collision between alkali-metal atoms the dominant interaction
leading to relaxation is the so-called spin-axis interaction
\cite{appelt_pra,spin_axis}, described by a hamiltonian of the
form ${\cal H}_{\rm sa }=\lambda
(3\sigma_{\zeta}\sigma_{\zeta}-1)$, where $\sigma_{\zeta}$ is the
projection of the total electron spin of the colliding atoms on
the inter-nuclear axis and $\lambda$ is a coupling constant. This
represents correlated decoherence \cite{preskill}, the effect of
which is that it preserves spin correlations. To prove that, we
consider two spin-1/2 particles in the triplet sub-space
\cite{kitagawa} state
$|\psi\rangle=\alpha|00\rangle+\beta(|01\rangle+|10\rangle)/\sqrt{2}+\gamma|11\rangle$
colliding along some axis $\mbox{\boldmath${\hat{\eta}}$}$. The
spin-axis hamiltonian reduces to ${\cal H}_{\rm sa }\sim\lambda
s_{\eta}^{(1)}s_{\eta}^{(2)}$. The change in the initial density
matrix $\rho=|\psi\rangle\langle\psi|$ induced by ${\cal H}_{\rm
sa }$ will be $d\rho=-i[{\cal H}_{\rm sa },\rho]dt$. We then
calculate the correlation $C_{yy}^{12}$ (or concurrence
\cite{conc}) in the state $\rho$, and find
$C_{yy}^{12}=(\beta^{2}-2\alpha\gamma)/4$. Whereas the change in
the expectation value $\langle S_{y}\rangle=\langle
s_{y}^{(1)}+s_{y}^{(2)}\rangle$ is found to be proportional to
$\lambda dt$, the change in the correlation $dC_{yy}^{12}$ is
found to be proportional to $(\lambda dt)^{2}$. This means that
 ${\cal H}_{\rm sa }$ induces dissipation, as expected, but preserves two-body quantum correlations of the spins.
 \newline\indent We now turn to the many-particle spin-squeezed state introduced in \cite{kitagawa},
 $|\Psi_{ss}\rangle=\sum_{l=0}^{N}{c_{l}|\phi_{l}^{\rm perm}\rangle}$,
 where $|\phi_{l}^{\rm perm}\rangle=\sum_{\rm perm}{|1^{\otimes
 l}0^{\otimes N-l}\rangle}$ is a permutation symmetric state with $l$ 1's
 and $N-l$ 0's, $c_{l}=i^{l}b_{l}$ and $b_{l}$ are given in
 \cite{kitagawa}. We take $dt$ to be the duration during which
 particles 1 and 2 have interacted through ${\cal H}_{\rm sa}$,
 resulting in a state change $|d\Psi_{ss}\rangle$.
 The change in the ensemble variance is
 $d(\Delta S_{y})^{2}=\langle d\Psi_{ss}|S_{y}^{2}|\Psi_{ss}\rangle+\langle\Psi_{ss}|S_{y}^{2}|d\Psi_{ss}\rangle$.
 Since $\langle S_{y}\rangle\approx 0$, it follows that
 $S_{y}|\Psi_{ss}\rangle$ is almost perpendicular to $|\Psi_{ss}\rangle$
 and therefore $S_{y}^{2}|\Psi_{ss}\rangle$ is proportional to
 $|\Psi_{ss}\rangle$. Thus $d(\Delta S_{y})^{2}\sim\mathfrak{Re}\{\langle
 d\Psi_{ss}|\Psi_{ss}\rangle\}$. By use of their symmetry, it is easily seen that the change
 induced by ${\cal H}_{\rm sa}$ in the states $|\phi_{l}^{\rm
 perm}\rangle$ is
\beq
 |\delta\phi_{l}\rangle=-i\lambda dt
 \left(\alpha|\chi_{l-2}\rangle+\beta|\chi_{l}\rangle+\alpha|\chi_{l+2}\rangle\right)
\eeq where the real coefficients $\alpha$ and $\beta$ depend on
the particular collision trajectory and the states
$|\chi_{m}\rangle$ contain a subset of the terms of the
corresponding states $|\phi_{m}^{\rm perm}\rangle$, with $m=l,l\pm
2$, and hence $\langle\chi_{m}|\phi_{m}^{\rm perm}\rangle$ is a
real number. Thus the overlap $\langle
 d\Psi_{ss}|\Psi_{ss}\rangle\sim i\lambda dt (\alpha
 c_{l-2}+\beta c_{l}+\alpha c_{l+2})c_{l}^{*}$, and therefore,
 since $b_{l}\approx b_{l\pm 2}$ for large $l$,
$\mathfrak{Re}\{\langle
 d\Psi_{ss}|\Psi_{ss}\rangle\}=0$.
 We have thus proved in the most general way that correlated
 relaxation preserves the ensemble variance $(\Delta S_{y})^{2}$,
 i.e. $d(\Delta S_{y})^{2}=0$ to first order in $dt$.
\newline\indent
On the contrary, independently acting decoherence processes will
tend to reduce the ensemble correlations during the measurement.
Based on the above considerations, when an atom
 decoheres independently, the change in $|\phi_{l}^{\rm
 perm}\rangle$ is proportional to a linear combination of $|\chi_{l\pm 1}^{\rm
 perm}\rangle$, and that leads to a non-zero real part of the
 overlap $\langle d\Psi_{ss}|\Psi_{ss}\rangle$, which is of order
 $dt$. The consequence of the decay of $(\Delta S_{y})^{2}$ can
 be simply described with an effective squeezing parameter
$\xi'>\xi$. Indeed, if we assume that there are two decoherence
mechanisms, one acting independently on each atom and one
preserving $(\Delta S_{y})^{2}$, with respective rates $1/T_{\rm
nc}$ and $1/T_{\rm c}$ (obviously $1/T_{2}=1/T_{\rm nc}+1/T_{\rm
c}$), using Eqs. (\ref{eq:spin_fl}) and (\ref{eq:corr}) we again
arrive at (\ref{eq:reduced_spin_fluct}), but with $\xi$ replaced
by \beq \xi'=\sqrt{{\xi^{2}+T_{\rm c}/T_{\rm nc}}\over {1+T_{\rm
c}/T_{\rm nc}}} \eeq If $T_{\rm c}/T_{\rm nc}\gg 1$, we get
$\xi'\approx 1$ and recover the case of uncorrelated spin noise
described by (\ref{eq:total_spin_fl}), with $T_{2}=T_{\rm nc}$.
This limit corresponds to the results obtained in
\cite{budker_qnd,ekert,kitagawa} where entanglement in the
presence of decoherence is shown not to offer any increase in
measurement precision beyond the uncorrelated ensemble case. On
the other hand, since there will always be an independently acting
relaxation mechanism, such as atom collisions with buffer gas
atoms or container walls, in the limit that $T_{\rm c}/T_{\rm
nc}\ll 1$ there is no point in attempting to reduce $\xi^{2}$
below this ratio. For example, using the spin-destruction cross
sections \cite{romalis2002} for K-K and K-He collisions, we find
that for a potassium density of [K]=$10^{15}~{\rm cm^{-3}}$ in the
presence of 1 atm of helium buffer gas, it is $T_{\rm c}\approx 1$
ms and $T_{\rm nc}/T_{\rm c}\approx 50$, suggesting, were it
experimentally feasible, the need for a squeezing parameter no
smaller than $\xi\approx 0.1$. The above considerations imply a
new fundamental sensitivity limit for  a frequency measurement.
Based on (\ref{eq:domega}) and assuming $\xi\approx\sqrt{T_{\rm
c}/T_{\rm nc}}$ we find \beq
\delta\omega_{L}={1\over{\sqrt{NT_{\rm
nc}T}}},\label{eq:new_limit} \eeq but now $T_{2}=T_{\rm c}\ll
T_{\rm nc}$. This is the second main result of this work.
Essentially, with spin squeezing we manage to suppress the effect
of correlated spin relaxation on the measurement precision.
Undoubtedly, the measurement precision still scales as
$1/\sqrt{N}$. However, $N$ can now be made as large as is
practically possible, since $T_{\rm nc}$ is atom-density
independent. Therefore, $\delta\omega_{L}$ can be made arbitrarily
small, in contrast to the uncorrelated-atoms case, where
$\delta\omega_{L}$ saturates at high densities since the
relaxation rate $1/T_{\rm c}$ is proportional to the atom density.
Furthermore $1/T_{\rm nc}$ represents relaxation due to
"technical" imperfections, such as relaxing cell walls, that is
not in any fundamental way prevented from reaching small values.
\newline\indent Since is is straightforward to show that
independently acting decoherence tends to reduce $(\Delta S_{y})^{2}$ to
the uncorrelated-atoms value of $N/4$ at a rate $1/T_{\rm nc}$, we
can more formally arrive at Eq. (\ref{eq:new_limit}) by solving the set of
equations
\begin{align}
&d\langle S_{y}\rangle =\omega_{L}\langle S_{x}\rangle_{0}e^{-t/T_{2}}dt-{{dt}\over {2T_{2}}}\langle S_{y}\rangle+\Delta S_{y}{{d\eta}\over\sqrt{T_{2}}}\\
&d(\Delta S_{y})^{2}=-\left((\Delta S_{y})^{2}-{N\over
4}\right){{dt}\over T_{\rm nc}}
\end{align}
where the decay of $\langle S_{x}\rangle$ has also been included
for completeness and initially $\Delta S_{y}(t=0)=\xi\sqrt{N/4}$.
For a measurement time $T_{\rm c}$, the above equations lead to a
sensitivity $\delta\omega_{L}=1/\sqrt{NT_{\rm c}T_{\rm nc}}$.
Averaging $n=T/T_{\rm c}$ such measurements leads again to
(\ref{eq:new_limit}).
\newline\indent
In conclusion, we have shown that by suppressing spin noise due to
correlated spin relaxation, spin squeezing results in an increased
measurement precision during a long measurement time in a
collision-dominated alkali-metal vapor. Similar comments apply to
the case of atomic clocks employing thermal alkali-metal vapors.
The spin-exchange interaction during binary collisions, which is
the main decoherence mechanism in these clocks, and of the form
${\cal H}_{se}\sim\mathbf{s}_{1}\cdot\mathbf{s}_{2}$,
 shares the same property with the spin-axis interaction, i.e. it results in correlated decoherence. Spin squeezing the clock transition could
 thus significantly boost the clock performance of frequency
 standards employing thermal alkali-metal vapors.

\end{document}